\documentclass[10pt, letter, twocolumn]{IEEEtran}
\usepackage{hyperref}
\usepackage{cleveref}
\usepackage[dvips]{color}
\usepackage{epsf}
\usepackage{times}
\usepackage{epsfig}
\usepackage{graphicx}
\usepackage{epstopdf}
\usepackage{algorithm}
\usepackage{algorithmic}
\usepackage{amsmath}
\usepackage{amssymb}
\usepackage{amsxtra}
\usepackage{multirow}
\usepackage{mathtools}
\usepackage{caption}

\usepackage{url}
\usepackage{subfigure}

\usepackage{mathtools,amsbsy,bbm,color}

\AtBeginDocument{%
 \abovedisplayskip = 0.04in
 \abovedisplayshortskip = 0.04in
 \belowdisplayskip =  0.04in
 \belowdisplayshortskip = 0.04in
 \topsep = 0.04in
 \textfloatsep = 0.05in
 \textheight = 9.35in
}
\begin{document}
%

\title{A Stochastic Geometry-based Demand Response Management Framework for Cellular Networks Powered by Smart Grid}


\author{ \IEEEauthorblockN{\large Muhammad Junaid Farooq, Hakim Ghazzai, and Abdullah Kadri} \\ \IEEEauthorblockA{Qatar Mobility Innovations Center (QMIC), Qatar Science \& Technology Park, Doha, Qatar},\\ Email: \{junaidf, hakimg, abdullahk\}@qmic.com. \vspace{-0.6cm}
{\thanks {\vspace{-0.4cm}\hrule \vspace{0.05cm} \indent This work was made possible by NPRP grant \# 6-001-2-001 from the Qatar National Research Fund (A member of The Qatar Foundation). The statements made herein are solely the responsibility of the authors.}}}

\maketitle

\begin{abstract}
In this paper, the production decisions across multiple energy suppliers in smart grid, powering cellular networks are investigated. The suppliers are characterized by different offered prices and pollutant emissions levels. The challenge is to decide the amount of energy provided by each supplier to each of the operators such that their profitability is maximized while respecting the maximum tolerated level of CO$_{2}$ emissions. The cellular operators are characterized by their offered quality of service (QoS) to the subscribers and the number of users that determines their energy requirements. Stochastic geometry is used to determine the average power needed to achieve the target probability of coverage for each operator. The total average power requirements of all networks are fed to an optimization framework to find the optimal amount of energy to be provided from each supplier to the operators. The generalized alpha-fair utility function is used to avoid production bias among the suppliers based on profitability of generation. Results illustrate the production behavior of the energy suppliers versus QoS level, cost of energy, capacity of generation, and level of fairness.\vspace{-0.05cm}
\end{abstract}

\vspace{-0.3cm}

\section{Introduction}
Cellular networks have seen an enormous growth in the number of connected subscribers, particularly after the advent of smart phones and 3G/4G mobile broadband technologies. To cater for the existing, as well as future communication requirements of emerging wireless technologies such as the 5G and internet of things (IoT), there is a need for massive network deployment, i.e., base stations (BSs) to increase network capacity of cellular operators~\cite{Ericsson_white_paper}. This will lead to a surge in the energy consumption of cellular networks and hence, negative environmental impact in the form of pollutant gases emissions mainly carbon dioxide (CO$_{2}$). It is estimated that the carbon footprint of wireless communications could almost triple in 2020 if no actions are taken~\cite{ericsson_report}. In order to counter the increasing carbon footprint of cellular networks, steps must be taken to limit the pollutant emissions by incorporating green renewable sources of energy alongside conventional fossil fuel energy generation. Moreover, effective energy management techniques are required in order to achieve the goal of reducing CO$_2$ emissions.

The development of the smart grid, a modern electricity grid that allows integration of energy generated from different sources, has enabled intelligent energy management both at the demand and supply side. The smart grid comprises of multiple energy suppliers, some of which provide renewable energy. The demand side management (DSM) in smart grid powered cellular networks allows operators to make profitable and environment-friendly energy procurement decisions according to the changes in price or carbon footprint of the energy \cite{Junaid_procurement}. On the other hand, an increase in energy demand from the side of cellular operators requires a decision at the smart grid level about the quantity of energy provided by the suppliers to the cellular operators and the price of energy. This is also known as demand response management (DRM)~\cite{demand_response}.

The objective of DRM is to determine the average amount of energy to be produced by each supplier in smart grid such that the profit of the suppliers is maximized while respecting the CO$_{2}$ emissions threshold set by the regulatory authority. For this, an accurate estimate of the total energy requirements of the cellular operators is required, that depends on the quality of their transmissions and the number of subscribers using their services. The average radiated power of BSs of each operator is the most crucial component in determining the total power consumption. Stochastic Geometry (SG)~\cite{martin_book} is employed to estimate the transmit power required by operators to provide their subscribers with a certain quality of service (QoS) level. SG allows the computation of important network statistics such as the probability of coverage (or outage) as spatial averages~\cite{jeffrey_coverage}. Since cellular networks involve BSs and users which can be modeled as points from a particular point process, we can use SG for performance analysis of such networks. The Poisson point process (PPP), which is well known for its analytical tractability, is used to model the location of operator BSs and subscribers. The resulting analysis enables us to ensure that, on average, any user in the network is served with a predefined QoS and to determine the corresponding power requirements.

Existing works on DRM in smart grid deal with capacity planning by a single energy supplier in response to the changes in the energy demand of generic consumers~\cite{demand_response}. Real-time pricing in smart grid with demand of the consumers is investigated in~\cite{drm2}. Most of the previous works are based on DRM for general electricity consumers. Few works have dealt with smart grid DRM for cellular networks. However, they are based on instantaneous network statistics~\cite{Hakim_drm,drm_cellular} without considering fairness in the supply decision. In this paper, a generalized $\alpha$-fairness-based DRM framework for cellular networks is proposed using average network statistics provided by SG. The system model comprises a set of cellular operators, characterized by the offered QoS to their subscribers, powered by a common pool of energy suppliers. SG is exploited to evaluate the total power requirements of the cellular networks. Then, closed-form expressions for the amount of energy provided by each supplier to the operators are derived while respecting the maximum tolerated CO$_{2}$ emissions level set by the environmental regulator. A generalized $\alpha$-fair utility function based on the profit of energy suppliers is used to avoid production bias towards any single supplier. The optimization is performed while taking into account the production capacity of each supplier, their cost of generation, the dynamic pricing strategy, and the pollutant emissions levels. Simulation results investigate the energy suppliers' behaviors according to different system parameters. Also, the obtained results show that suppliers' production decisions are adapted to the changing demand of cellular networks to achieve maximum profitability in an environment-friendly manner.

%
\vspace{-0.15cm}
\section{System Model}
\label{sec2}
\subsection{Network Geometry}
The network consists of $N_{\text{op}}$ mobile operators serving their subscribers in an area of $\mathcal{A}$ km$^{2}$. The BSs of each mobile operator $l=1,\ldots,N_{\text{op}}$ are randomly deployed in $\mathbb{R}^{2}$ according to a homogeneous PPP denoted by $\Phi^{(l)}$ with density $\lambda_{\text{BS}}^{(l)}$ BS/km$^2$. The total number of BSs that belong to each operator is $N_{\text{BS}}^{(l)} = \lambda_{\text{BS}}^{(l)} \mathcal{A}$. Each mobile operator uses frequency reuse within its network to avoid adjacent cell interference. The set of BSs using the same frequency are, therefore, separated by a certain \emph{exclusion distance} denoted by $R$. The location of the interfering BSs of each operator are modeled by the Mat\'ern hard core point process (MHCPP) obtained via a dependent thinning of the original PPP. No two selected BSs in the MHCPP are closer than the exclusion distance. This new set is denoted by $\tilde{\Phi}^{(l)} \subseteq \Phi^{(l)}$ (see \cite{hesham_survey} for details). For tractability of analysis, the MHCPP is approximated by an equi-dense PPP. The intensity of the MHCPP is evaluated as $\Lambda_{\text{BS}}^{(l)} = \frac{1-e^{-\lambda_{\text{BS}}^{(l)} \pi R^{2}}}{\pi R^{2}},$~\cite{stoyan_book}. The subscribers of operator $l$ are uniformly distributed in $\mathbb{R}^{2}$ with density $\lambda_{u}^{(l)}$ users/km$^{2}$. Each user is associated to its nearest BS and, hence, the coverage regions of BSs form a Voronoi tessellation~\cite{jeffrey_coverage}. The number of users served by each BS of operator $l$ are evaluated as $\mathcal{N}_{u}^{(l)} = \lambda_{u}^{(l)}\mathcal{A} / N_{\text{BS}}^{(l)}$.

\vspace{-0.08in}
\subsection{Operator Characteristics}
In this framework, each mobile operator is characterized by the offered QoS, determined by the downlink signal-to-interference-plus-noise ratio (SINR) thresholds $T^{(l)}$ and the probability of coverage $\mathcal{P}^{(l)}$ for $l= 1,\ldots, N_{\text{op}}$. The assumption is that a user connected to the BS of operator $l$, is served only if the downlink SINR at the user exceeds the pre-defined threshold $T^{(l)}$. To ensure that each operator meets the set transmission quality standards,
the transmit power levels of its BSs need to be adjusted. The total transmit power from all the BSs determines the energy requirements of the operator.
\vspace{-0.08in}
\subsection{Channel Model}
It is assumed that the BSs of operator $l$ transmit with an average power denoted by $P_{t}^{(l)}$ for $l = 1,\ldots,N_{\text{op}}$ to serve a single user. The power decays with the distance according to the power law and, hence, the received power at a test user located at the origin from a BS located at $x_{j}$ is given as, $P_{r}^{(l)} = P_{t}^{(l)}h_{j}^{(l)} \|x_{j}^{(l)}\|^{-\eta}$, where $h_{j}^{(l)}$ represents the channel fast fading effect and is assumed to be independent and identically distributed (i.i.d.) exponential random variables with mean $\mu^{-1}$ for each $j,l$, $\|.\|$ represents the vector norm, and $\eta$ is the path-loss exponent. Without loss of generality, it is assumed that the effects of shadowing are catered for by the randomness of the BS locations. The downlink SINR at a typical user located at the origin served by operator $l$ can be expressed as:
\begin{equation}
\small
\textrm{SINR}^{(l)} = \frac{P_{t}^{(l)} h_{0}^{(l)} {r^{(l)}}^{-\eta}}{\sigma^{2} + \sum \limits_{j \in \tilde{\Phi}^{(l)} \backslash 0} P_{t}^{(l)} h_{j}^{(l)} \| x_{j}^{(l)}\|^{-\eta}},
\end{equation}
\normalsize
where $h_{0}^{(l)}$ and $r^{(l)}$ are the channel fading and the distance between the test user and the nearest BS of operator $l$, respectively, and $\sigma^{2}$ is the Gaussian noise power. The interference power denoted by $\mathcal{I}_{\text{agg}}^{(l)}$ is $\sum \limits_{j \in \tilde{\Phi}^{(l)} \backslash 0} P_{t}^{(l)} h_{j}^{(l)} \|x_{j}^{(l)}\|^{-\eta}$, where $\tilde{\Phi}^{(l)} \backslash 0$ denotes the set of BSs of operator $l$ excluding the BS closest to the considered user.

\vspace{-0.1in}
\subsection{Power Consumption Model for Mobile Operators}
Each BS of operator $l$ is considered to be equipped with a single omni-directional antenna. Its average power consumption, denoted by $P_{\text{BS}}^{(l)}$, is expressed as follows~\cite{TVTjournal1}:
\begin{equation}
\small
P_{\text{BS}}^{(l)} = a\,P_{\text{tx}}^{(l)}+b,
\end{equation}
\normalsize
where $P_{\text{tx}}^{(l)}$ is the average power radiated by each BS of operator $l$. The coefficient $a$ corresponds to the power consumption that scales with the radiated power due to amplifier and feeder losses. The term $b$ models an offset of BS site power which is consumed independently of the average transmit power and is due to signal processing, battery backup, and cooling. $P_{\text{tx}}^{(l)}$ can be evaluated as $P_{\text{tx}}^{(l)} =  P_{t}^{(l)} \mathcal{N}_{u}^{(l)}$.
\vspace{-0.1in}
\subsection{Energy Pricing and Environmental Impact}
In this study, it is assumed that the cellular networks are powered by a smart grid where $N_{\textrm{R}}$ different sources exist to supply energy with different prices and have different pollutant levels depending on the nature of the generated energy. The suppliers use a dynamic pricing strategy, whereby, the unit charge of supplied energy depends directly on the energy demand. In order to increase production, an energy supplier needs to employ additional infrastructure and resources, thus bearing additional expenses. Therefore, the price of energy from supplier $n$ charged to operator $l$, denoted by $\pi^{(n,l)}$, is expressed in monetary units (MU) as follows:
\begin{equation}
\label{dynamic_pricing}
\small
\pi^{(n,l)} = w^{(n)} \left(  \frac{q^{(n,l)}}{\bar{Q}^{(n)}} \right)^{\gamma^{(n)}},
\end{equation}
\normalsize
where $w^{(n)}$ is the benchmark price charged per unit by supplier $n$, $q^{(n,l)}$ is the average energy provided by supplier $n$ to operator $l$, $\bar{Q}^{(n)}$ is the maximum generation capacity of each supplier, and $\gamma^{(n)}$ is a non-negative integer representing the price sensitivity of supplier $n$. If $\gamma^{(n)} = 0$, the dynamic pricing is ignored and the suppliers charge a fixed price for energy independently of the quantity purchased. Otherwise, the price increases with the demand and vice versa.

The suppliers have different environmental impacts in terms of CO$_2$ emissions depending on the quantity and type of energy generated. This can be quantified using the following pollutant emissions function $\mathcal{F}$ (see \cite{TVTjournal1}):
\begin{equation}
\small
{\mathcal F}(q^{(n,l)})=\psi_n \sum \limits_{l=1}^{N_{\text{op}}}{\left(q^{(n,l)} \right)}^2+\phi_n \sum \limits_{l=1}^{N_{\text{op}}} q^{(n,l)},
\end{equation}
\normalsize
where $\psi_{n}$ and $\phi_{n}$ are the pollutant emissions coefficients of supplier $n$.
\vspace{-0.12in}
\section{Methodology of Analysis}\label{sec3}
In this section, the probability of coverage for a typical user in the network of operator $l$ is determined using analytical tools from SG. This helps us in identifying the transmit power requirements of each operator to meet the transmission quality standards and, hence, determining the average power consumption of the operators. Consequently, an optimization framework is developed to supply the required energy from the smart grid in an environment-friendly yet profitable manner.
\vspace{-0.12in}
\subsection{Operator Coverage Analysis}
SG is employed to evaluate the probability that a typical user of operator $l$ is covered by the cellular network. The coverage is defined as the probability that the received SINR per user exceeds a target service threshold $T^{(l)}$. In other words, it corresponds to the complementary cumulative distribution function (ccdf) of the SINR, denoted by $\mathcal{P}^{(l)}$, evaluated as $\mathcal{P}^{(l)} = \mathbb{P}[\text{SINR}^{(l)} > T^{(l)}]$ and it can be expressed as~\cite{jeffrey_coverage}:
\begin{align}
\small
&\mathcal{P}^{(l)}(P_{t}^{(l)},T^{(l)}) = \int \limits_{0}^{\infty} 2 \pi r \lambda_{\text{BS}}^{(l)} e^{- \lambda_{\text{BS}}^{(l)} \pi r^{2}} e^{-\frac{\mu T^{(l)} \sigma^{2} r^{\eta}}{P_{t}^{(l)}}} \times \nonumber \\
& \qquad \qquad \qquad \qquad \left( e^{- \Lambda_{\text{BS}}^{(l)} \pi {T^{(l)}}^{2/\eta} r^{2} \int \limits_{{T^{(l)}}^{-2/\eta}}^{\infty} \frac{1}{1+u^{\eta/2}} du } \right) dr.
\label{coverage_eq}
\end{align}
\normalsize
Note that the probability of coverage depends on the transmitted power and the target SINR threshold of each operator. Therefore, for a fixed $\mathcal{P}^{(l)}$, the transmit power can be compromised to achieve a higher SINR threshold and vice versa.
\vspace{-0.3in}
\subsection{Problem Formulation and Solution}
The decision on production quantity by the suppliers is based on their profitability as well as the CO$_2$ emissions. The profitability of the suppliers depends on the quantity of energy supplied to mobile operators $\mathbf{Q}$, where $\mathbf{Q}$ is a matrix of size $N_{\text{R}} \times N_{\text{op}}$ with elements $q^{(n,l)} \ \forall \ n=1,\ldots,N_{\text{R}}, l = 1,\ldots,N_{\text{op}}$, the per unit price of energy set by suppliers $\pi^{(n,l)}$, and the per unit cost of generation $c^{(n)}$. The profit of supplier $n$ can be expressed as follows:
\begin{equation}\label{profit_utility}\small
  \Pi_{n}(\mathbf{Q}) = \sum \limits_{l=1}^{N_{\text{op}}} q^{(n,l)} \left(  \pi^{(n,l)} - c^{(n)}\right), \ \forall \ n =1,\ldots,N_{R}.
\end{equation}
\normalsize

The environment friendliness of the suppliers is determined by the pollutant emissions levels that depend on the amount of energy supplied. It is quantified by the cost function $
\mathcal{C}(\mathbf{Q}) =  \sum \limits_{n = 1}^{N_{\textrm{R}}}{\mathcal F}(q^{(n,l)})$. In order to reduce pollutant emissions, the smart grid needs to produce more green energy from renewable sources while curtailing the use of fossil fuel sources. However, the renewable energy generation might be expensive and/or limited in amount and therefore, may lead to lower profitability for the suppliers. Hence, the suppliers use multiple energy sources depending on their cost, available amount, and environmental impact in order to increase their revenues while respecting the maximum tolerated CO$_2$ emissions level imposed by the regulator.

The objective now is to solve a constrained optimization problem which results in an optimal matrix $\mathbf{Q}^{*}$ that maximizes a utility $\mathcal{U}$ based on the vector $\mathbf{\Pi} = [\Pi_{1}, \Pi_{2} , \ldots, \Pi_{N_{R}}]$ containing the individual profits of the energy suppliers. The optimization problem is expressed as follows:
\begin{align}
\small
& \underset{\mathbf{Q}}{\text{maximize}}\quad \mathcal{U}(\mathbf{\Pi}) \label{opt_problem} \\
& \text{subject to}\quad\sum \limits_{l=1}^{N_{\text{op}}} N_{\text{BS}}^{(l)} q^{(n,l)} \leq \bar{Q}_{n}, \; \forall\; n = 1, \ldots, N_{\text{R}}, \label{max_energy} \\
&\sum \limits_{n=1}^{N_{\text{R}}} q^{(n,l)} = N_{\text{BS}}^{(l)}(a \  P_{t}^{(l)} N_{\text{u}}^{(l)} + b)\tau, \hfill{\forall\; l = 1, \ldots, N_{\text{op}}}, \label{consumption} \\
& \mathcal{C}(\mathbf{Q}) \leq \mathcal{C}_{\text{th}}.  \label{CO2}
\end{align}
\normalsize
The constraint \eqref{max_energy} enforces the total amount of energy produced by a supplier $n$ not to exceed its maximum energy production capacity $\bar{Q}_{n}$. Assuming that there are no losses, the average energy provided by the suppliers should be equal to the total power consumption of all BSs of operator $l$ during its operation time $\tau$, as indicated in constraint \eqref{consumption}. The total transmit power for each operator $P_{t}^{(l)}$ can be obtained by solving \eqref{coverage_eq} for a fixed QoS requirement $\mathcal{P}^{(l)}$. As the expression of $\mathcal{P}^{(l)}$ involves computation of a complicated integral which does not have a closed form, it is solved numerically using Newton's method to obtain $P_{t}^{(l)}$, $\forall l$. In constraint \eqref{CO2}, a restriction is imposed on the pollutant emissions due to energy consumption. This forces the suppliers to produce more green energy to remain within regulated limits, denoted by $\mathcal{C}_{\text{th}}$.
\vspace{-0.15in}
\subsection{Utility Function}
Since the objective is to determine the optimal amount of energy provided by the suppliers in the smart grid to multiple mobile operators, a fair power sharing framework needs to be introduced. Therefore, a unified mathematical formulation for power allocation known as $\alpha$-fairness is used, where the parameter $\alpha$ determines the degree of fairness of profit distribution among the operators \cite{alpha_fair}. The generalized $\alpha$-fairness utility function of the supplier profits is defined as follows:
\begin{align}\small
\mathcal{U}(\mathbf{\Pi}) =
\left\{
	\begin{array}{ll}
		\sum \limits_{n=1}^{N_{\text{R}}} \frac{\Pi_{n}^{1-\alpha}}{1-\alpha},  & \mbox{if } \alpha \geq 0, \alpha \neq 1,\\
		\sum \limits_{n=1}^{N_{\text{R}}} \log(\Pi_{n}), & \mbox{if } \alpha = 1.
	\end{array}
\right.
\end{align}
\normalsize
The parameter $\alpha$ controls the tradeoff between the efficiency as well as the fairness in energy provided by the suppliers to the operators. In other words, the framework tends to avoid unfair allocation of energy depending on the value of $\alpha$. For example, situations where the suppliers only supply the most expensive energy to meet the demand in order to achieve maximum profit. In the special cases, when $\alpha = 0,1,\infty$, the utility reduces to the sum, proportional fair and max-min fair utility, respectively. These cases are described as follows:\\
\textbf{Sum Utility ($\alpha = 0$):} The utility of this metric is equivalent to the sum of the profits of the energy suppliers $(\mathcal{U}(\mathbf{\Pi}) = \sum \limits_{n=1}^{N_{\text{R}}}\Pi_{n})$. This approach promotes suppliers with higher profit margins by offering them the possibility to supply the most expensive energy to the operators at the least cost via smart grid to maximize their profits. On the other hand, suppliers with lower energy price will be deprived from the possibility to sell the cheapest energy as their profits are already low.\\
\textbf{Proportional fair utility ($\alpha = 1$):} The proportional fair metric maximizes the sum of the log of the profit of suppliers. It is equivalent to the maximization of the geometric mean of the profits $\Pi_{n}$ i.e. $(\mathcal{U}(\mathbf{\Pi}) = (\prod_{n=1}^{N_{\text{R}}} \Pi_{n})^{1/N_{\text{R}}})$, which is equivalent to $(\mathcal{U}(\mathbf{\Pi}) = \sum_{n=1}^{N_{\text{R}}} \ln (\Pi_{n}))$. The proportional fair (PF) metric is fairer since an energy supplier with a profit close to zero will make the entire utility go to zero. Hence, this metric avoids having a very low profit for any of the suppliers. In addition, it reasonably promotes suppliers with good profit margins since a high profit will contribute in increasing the product.\\
\textbf{Max-Min Utility ($\alpha \rightarrow \infty$):} Max-Min utilities are a family of utility functions attempting to maximize the minimum profit among all energy suppliers: $\mathcal{U}(\mathbf{\Pi}) = \underset{n}{\min}(\Pi_{n})$. By increasing the priority of suppliers having lower profit margins, Max-Min utilities lead to more fairness in the system. In order to simplify the problem for this approach, a new decision variable $\Pi_{\text{min}}=\underset{n}{\min}(\Pi_{n})$ is introduced. Therefore, the optimization problem becomes:
\begin{align}
\small
& \underset{\mathbf{Q},\Pi_{\text{min}}}{\text{maximize}}
& & \Pi_{\text{min}} \\
& \text{subject to}
& & \eqref{max_energy},\eqref{consumption} \ \text{and} \ \eqref{CO2} , \\
&&& \Pi_{n} \geq \Pi_{\text{min}}, \forall n=1,\ldots,N_{\text{R}}. \label{Pi_min_const}\vspace{-0.05in}
\end{align}
\normalsize
\vspace{-0.3in}
\subsection{Solution}
The optimization problem presented in \cref{opt_problem,max_energy,consumption,CO2} can be solved for special cases using the Lagrangian method~\cite{Boyd} exploiting its strong duality property as follows:
\begin{equation}\label{lagrangian}\small
   \underset{\boldsymbol{\delta} \geq 0}{\mathrm{minimize}} \quad \underset{\mathbf{Q} \geq 0}{\mathrm{maximize}} \qquad L(\mathbf{Q},\boldsymbol{\delta},\boldsymbol{\xi},\zeta),
\end{equation}
\normalsize
where $L(\mathbf{Q},\boldsymbol{\delta},\boldsymbol{\xi},\zeta)$ is the Lagrangian, $\boldsymbol{\delta} = [\delta_{1}, \ldots, \delta_{N_{\text{R}}}]$, $\boldsymbol{\xi} = [\xi_{1}, \ldots, \xi_{N_{\text{op}}}]$, and $\zeta$ are the Lagrange multipliers corresponding to the constraints in \eqref{max_energy}, \eqref{consumption}, and \eqref{CO2} respectively. Deriving the optimal amount of energy for a general value of $\alpha$ is highly elaborate and is left for future work. The closed form expressions for the special cases, i.e., $\alpha = 0,1,\infty$ are provided here. However, the investigation of the performance for different values of $\alpha$ is done via simulations in Section~IV.

For the case of Sum utility, i.e., $\alpha = 0$, the optimal amount of energy supplied by supplier $n$ to operator $l$, $q^{(n,l)*}$ is:
\normalsize
\begin{align}
\small
q^{(n,l)*} =
\left\{
	\begin{array}{ll}
    \hspace{-0.1in}\frac{1}{2 \zeta \psi_{n} N_{\text{BS}}^{(l)}} \left(  w^{(n)} - c^{(n)} - \delta_{n} N_{\text{BS}}^{(l)} - \zeta \phi_{n} N_{\text{BS}}^{(l)}  +  \xi_{l}  \right), \\
		\hspace{+0.7cm} \mbox{ if } \gamma_{n} = 0,   \\ 
    \hspace{-0.1in}\frac{1}{  2\left( \zeta \psi_{n} N_{\text{BS}}^{(l)} - \frac{w^{(n)}}{\bar{Q}_{n}} \right)} \left(  - c^{(n)} - \delta_{n} N_{\text{BS}}^{(l)} - \zeta \phi_{n} N_{\text{BS}}^{(l)}  +  \xi_{l}  \right),\\ \hspace{+0.7cm}\mbox{ if } \gamma_{n} = 1.
	\end{array}
\right.
\label{sum_utility_exp}
\end{align}
\normalsize
We can deduce that $q^{(n,l)*}$ is inversely proportional to the coefficient related to CO$_2$ emissions of energy supplier $n$ i.e., $\zeta$. The Sum utility depends on the prices charged by the suppliers where the higher the price, the higher the amount of energy supplied from that source.

For the case of Proportional fair utility, i.e., $\alpha = 1$, the optimal amount of energy supplied by supplier $n$ to operator $l$ is given as follows:
\normalsize
\begin{align}
q^{(n,l)*} =
\left\{
	\begin{array}{ll}
	\hspace{-0.1in} \frac{1}{2 \zeta \psi_{n} N_{\text{BS}}^{(l)}} \Big(  \left(w^{(n)} - c^{(n)} \right) \prod \limits_{\substack{m = 1 \\ m \neq n}}^{N_{\text{R}}} \Pi_{m} - \delta_{n} N_{\text{BS}}^{(l)} - \\ \zeta \phi_{n} N_{\text{BS}}^{(l)}  +  \xi_{l}  \Big), \mbox{if } \gamma_{n} = 0,   \\
	\hspace{-0.1in} \frac{1}{\frac{2 w^{(n)}}{\bar{Q}_{n}}  \prod \limits_{\substack{m = 1 \\ m \neq n}}^{N_{\text{R}}} \Pi_{m}   - 2 \zeta \psi_{n} N_{\text{BS}}^{(l)}} \Big(   c^{(n)} \prod \limits_{\substack{m = 1 \\ m \neq n}}^{N_{\text{R}}} \Pi_{m} + \delta_{n} N_{\text{BS}}^{(l)} + \\ \zeta \phi_{n} N_{\text{BS}}^{(l)}  -  \xi_{l}  \Big), \mbox{if } \gamma_{n} = 1.
	\end{array}
\right.
\end{align}
\normalsize
The supplied energy $q^{(n,l)*}$ depends directly on the product of other supplier profits i.e., $\Pi_{m}$. Hence, this approach avoids having any supplier with very low profit and maximizes the product of all suppliers’ profits simultaneously.

For the case of Max-Min fair utility, i.e., $\alpha \rightarrow \infty$, the optimal amount of energy supplied by supplier $n$ to operator $l$ is given as follows:
\normalsize
\begin{align}
\small
q^{(n,l)*} =
\left\{
	\begin{array}{ll}
	\hspace{-0.1in}\frac{1}{2 \zeta \psi_{n} N_{\text{BS}}^{(l)}} \left(  \theta_{n}( w^{(n)} - c^{(n)} ) - \delta_{n} N_{\text{BS}}^{(l)} - \zeta \phi_{n} N_{\text{BS}}^{(l)}  +  \xi_{l}  \right), \\
	\hspace{0.7cm}\mbox{if } \gamma_{n} = 0,  \\ 
	\hspace{-0.1in}\frac{1}{ 2 \left( \zeta \psi_{n} N_{\text{BS}}^{(l)} -  \frac{w^{(n)}}{\bar{Q}_{n}} \theta_{n} \right)   } \left(  c^{(n)} \theta_{n} + \delta_{n} N_{\text{BS}}^{(l)} + \zeta \phi_{n} N_{\text{BS}}^{(l)}  -  \xi_{l}  \right),\\
	\hspace{0.7cm}\mbox{if } \gamma_{n} = 1.
	\end{array}
\right.
\end{align}
\normalsize
$\theta_{n}$ are the additional Lagrange multipliers corresponding to \eqref{Pi_min_const}. By taking the derivative of the Lagrangian with respect to $\Pi_{\text{min}}$, we can deduce that $\sum_{n=1}^{N_{\text{R}}} \theta_{n} = 1$. By comparing with \eqref{sum_utility_exp}, we can notice that $\theta_{n}$ control the priority of the energy supply according to the profitability of the suppliers.

Note that if a fixed pricing strategy is used, i.e., $\gamma_{n} = 0$, the amount of energy supplied is directly proportional to the price of energy, whereas for dynamic pricing strategy, the amount of energy supplied is inversely related to the price of energy. This is because the energy price increases with demand under dynamic pricing policy. Hence, the amount supplied should decrease with increasing price.

The optimal Lagrange multipliers $\boldsymbol{\delta}^{*}$, $\boldsymbol{\xi}^{*}$, $\zeta^{*}$ and $\theta_{n}^{*}$ can be obtained using the subgradient method (see \cite{subgradient}) or other heuristic approaches. In order to achieve the optimal solution, we can start with random initial values of the Lagrange multipliers and evaluate the corresponding amount of energy. Then, we update the Lagrange multipliers at the next iteration with a step size updated according to the non-summable diminishing step length policy. The values of the optimal amount of energy supplied and the Lagrange multipliers are updated until convergence.
\vspace{-0.05in}
\section{Simulation Results}
\label{sec4}
A cellular network with $N_{\text{op}}=3$, denoted by Op. 1, Op. 2 and Op. 3, is considered. The BSs of each operator are distributed uniformly in $\mathbb{R}^{2}$ according to PPPs with BS density $\lambda_{\text{BS}}^{(1)} = \lambda_{\text{BS}}^{(2)} = \lambda_{\text{BS}}^{(3)} = (\pi(200)^{2})^{-1} \ \text{BS} / \textrm{m}^{2}$ respectively. The simulation area $\mathcal{A}$ is chosen to be a square region of side $10$ km$^2$ in $\mathbb{R}^{2}$. The subscribers of the operators are also uniformly distributed in the simulation area with intensity $\lambda_{u}^{(1)} = 15$ users/km$^{2}$, $\lambda_{u}^{(2)} = 30$ users/km$^{2}$, and $\lambda_{u}^{(3)} = 40$ users/km$^{2}$. The channel is assumed to experience exponential fading with a mean channel gain $\mu = 1$. The operators are differentiated by the offered SINR levels and the QoS. The SINR levels offered by the three operators are $T^{(1)} = 15$ dB, $T^{(2)} = 10$ dB, and $T^{(3)} = 5$ dB while the QoS requirements are specified as $\mathcal{P}^{(1)} = 0.7$, $\mathcal{P}^{(2)}= 0.8$, and $\mathcal{P}^{(3)} = 0.9$. Without loss of generality, the BS operation time $\tau$ is set to be $1$ s.

All BSs use the same power model \cite{TVTjournal1} with $a = 7.84$ and $b = 71.5$. The path loss exponent $\eta = 4$ and the noise power is $\sigma^{2} = -115$ dB. The BSs procure energy from a smart grid in which there are several available energy suppliers offering different prices and CO$_2$ emissions levels. It is assumed that there are three available energy suppliers (i.e., $N_{\text{R}} = 3$), namely; Sup. $1$, Sup. $2$, and Sup. $3$. They are characterized as follows: Sup. $1$ provides the cheapest energy at a benchmark price of $w^{(1)} = 1$ MU but with the highest carbon footprint, i.e., pollutant coefficients $\psi_{1} = 0.004$ and $\phi_{1} = 0.001$, Sup. $2$ provides energy with an intermediate benchmark price and lower carbon footprint ($w^{(2)} = 2 \ \textrm{MU}, \psi_{2} = 0.002$, $\phi_{2} = 0.0005$), and Sup. $3$ offers clean but expensive energy (i.e., $w^{(3)} = 3$ MU, $\psi_{3} = 0$, $\phi_{3} = 0.0001$). The maximum CO$_{2}$ emissions level is taken as $\mathcal{C}_{\text{th}}$ = $10^7$ kg/h. The costs of energy production for the suppliers is $c^{(1)} = 0.1$ MU, $c^{(2)} = 0.5$ MU, and $c^{(3)} = 2.5$ MU. The base profit margin of the suppliers is, therefore, $0.9$ MU, $1.5$ MU and $0.5$ MU, respectively. Each supplier has a maximum energy production capacity which is assumed to be equal (i.e., $\bar{Q}_{1} = \bar{Q}_{2} = \bar{Q}_{3} = 150$ kJ), unless otherwise stated.
\begin{figure}[t]
\begin{center}
\subfigure[Energy consumed.]{\label{Econsumed_T1_fig}\includegraphics[width=2.35in]{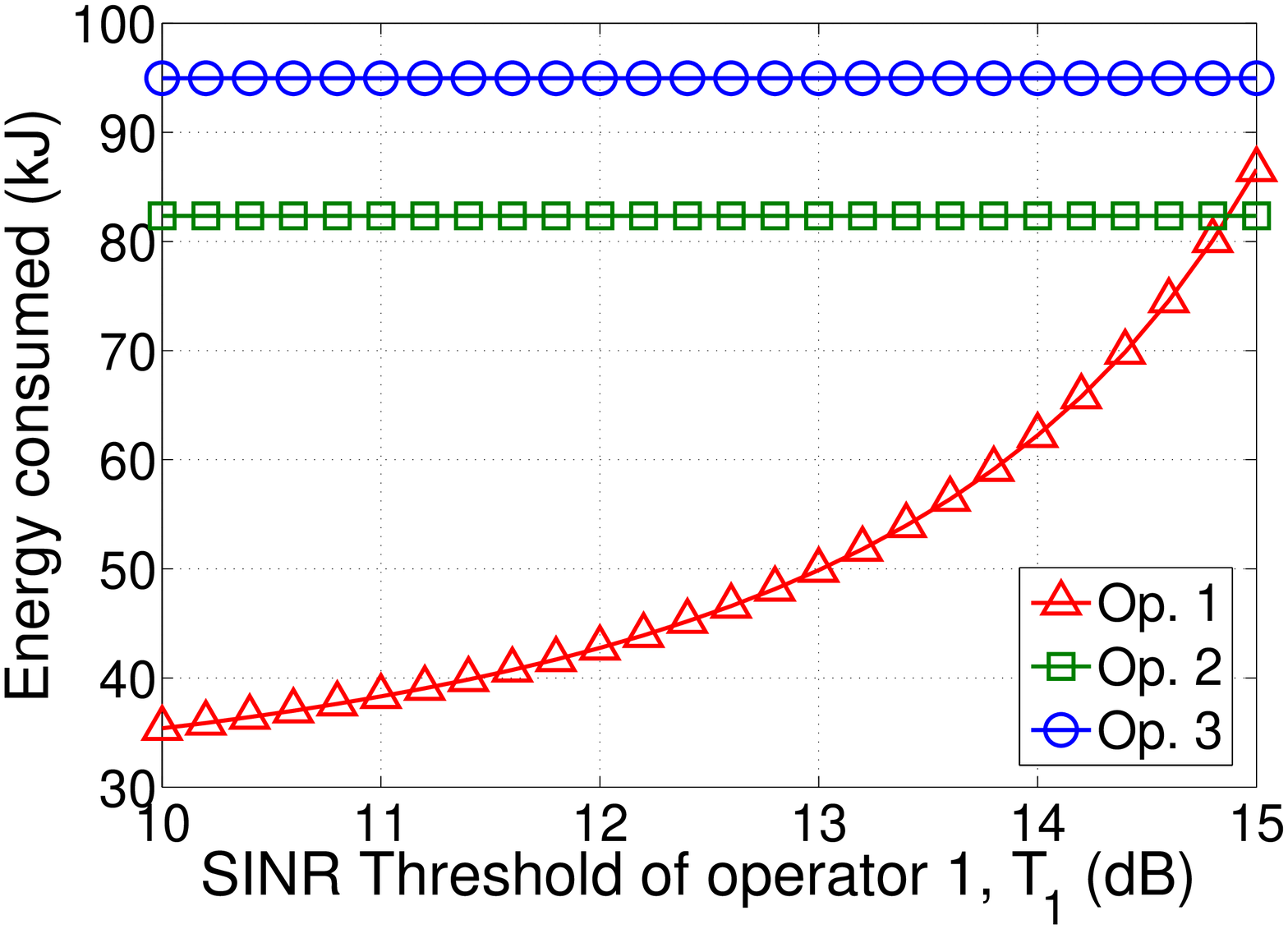}} \\
\vspace{-0.1in}
\subfigure[Price of energy.]{\label{price_vs_T1}\includegraphics[width=2.35in]{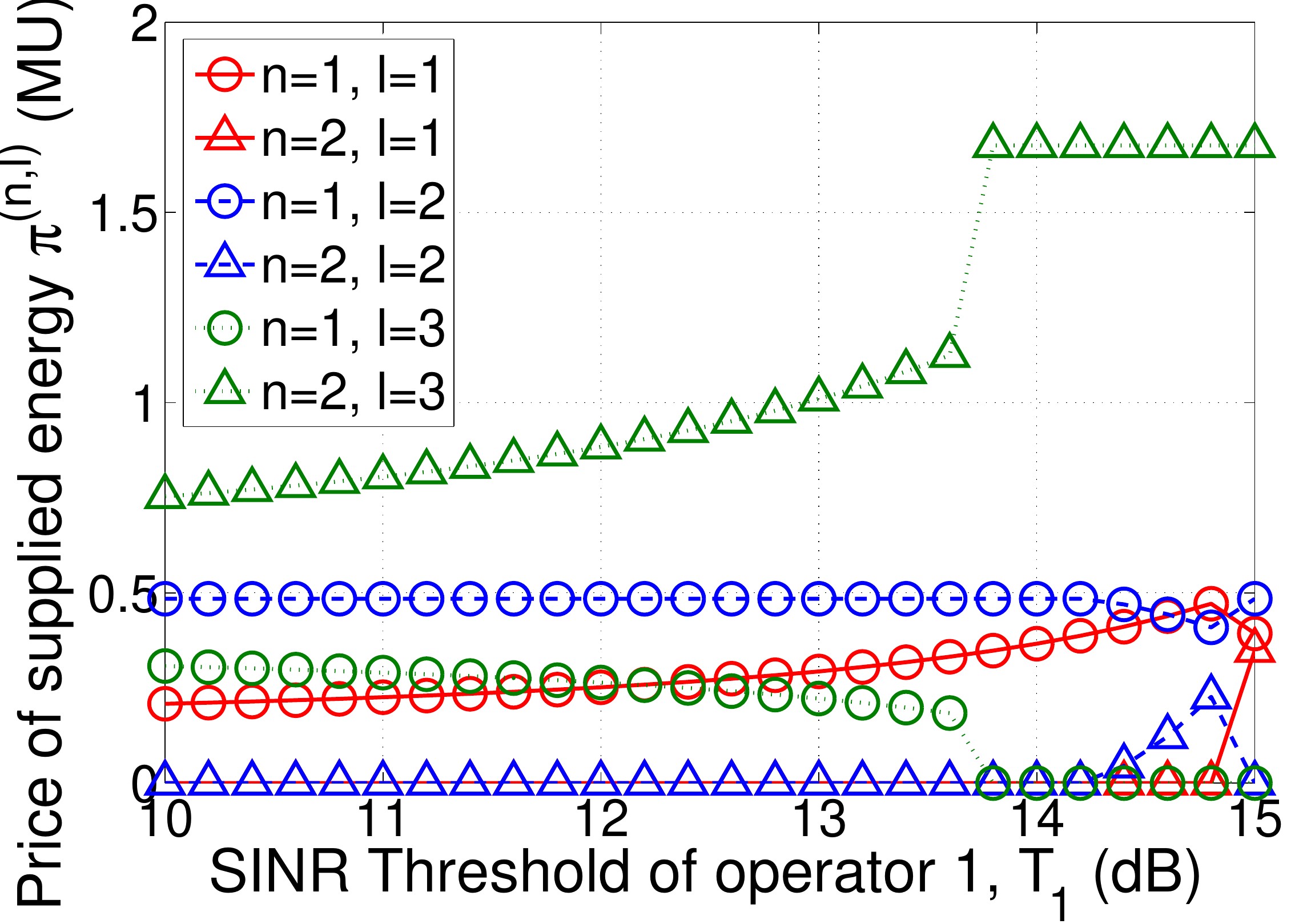}}
\end{center}
\vspace{-0.08in}
\caption{\small{Price of energy charged by suppliers to operators and the total energy consumed by operators versus SINR threshold of Op. 1. The simulation parameters are: $\alpha = 0$, $\gamma_{1}~=~\gamma_{2}~=~1$, $\bar{Q}_{1}~=~\bar{Q}_{2}$~=~170 kJ, $\mathcal{C}_{\text{th}}$ = $4.5 \times 10^{7}$ kg/h.}}
\label{fig1}
\end{figure}

The objective is to study the behavior of the system while changing system parameters and thresholds. Fig.~\ref{fig1} plots the price of energy charged to operators and the total energy consumed by the operators versus the SINR threshold of Op. 1. For ease of exposition, $N_{\text{R}} = 2$, i.e., Sup. 1 and Sup. 3, is used in Fig.~\ref{fig1}. We set $\gamma_{n}= 1$, which means that the dynamic pricing is in effect, and $\alpha = 0$, which means that the Sum utility is used. It can be observed from Fig.~\ref{Econsumed_T1_fig}, that an increase in transmission quality standards of a mobile operator increases its energy consumption due to additional transmit power required. This increased demand triggers a response from the suppliers as shown in Fig.~\ref{price_vs_T1}. Since Op. 3 is the most power hungry operator, it is provided with the most expensive energy followed by Op. 2 and Op. 1, respectively. Op. 1 and Op. 2 are completely powered by the cheapest energy supplier with high carbon footprint while Op. 3 is powered by both suppliers. As the power requirements of Op. 1 increase, its price also increases due to the dynamic pricing policy. To supply more energy to Op. 1 from Sup. 1, the framework progressively cuts down the energy supplied by Sup. 1 to Op. 3 and increases the energy supplied by Sup. 2 to Op. 3. As soon as CO$_2$ emissions threshold is reached, Op. 3 suffers further by completely being supplied by Op. 2 with very expensive energy. Further increase in demand is compensated by providing Op. 2 with more green expensive energy. Finally, the framework shows an abrupt change in the production decision after $T_{1} = 14.8$ dB. This is because Op. 1 becomes the most power hungry operator as evident from Fig.~\ref{Econsumed_T1_fig}. Therefore, it starts to receive green expensive energy and less cheap energy. The results in Fig.~\ref{fig1} motivate the introduction of fairness to the framework to avoid price exploitation of any single operator.

\begin{figure}[t]
  \centering
  \includegraphics[width=2.35in]{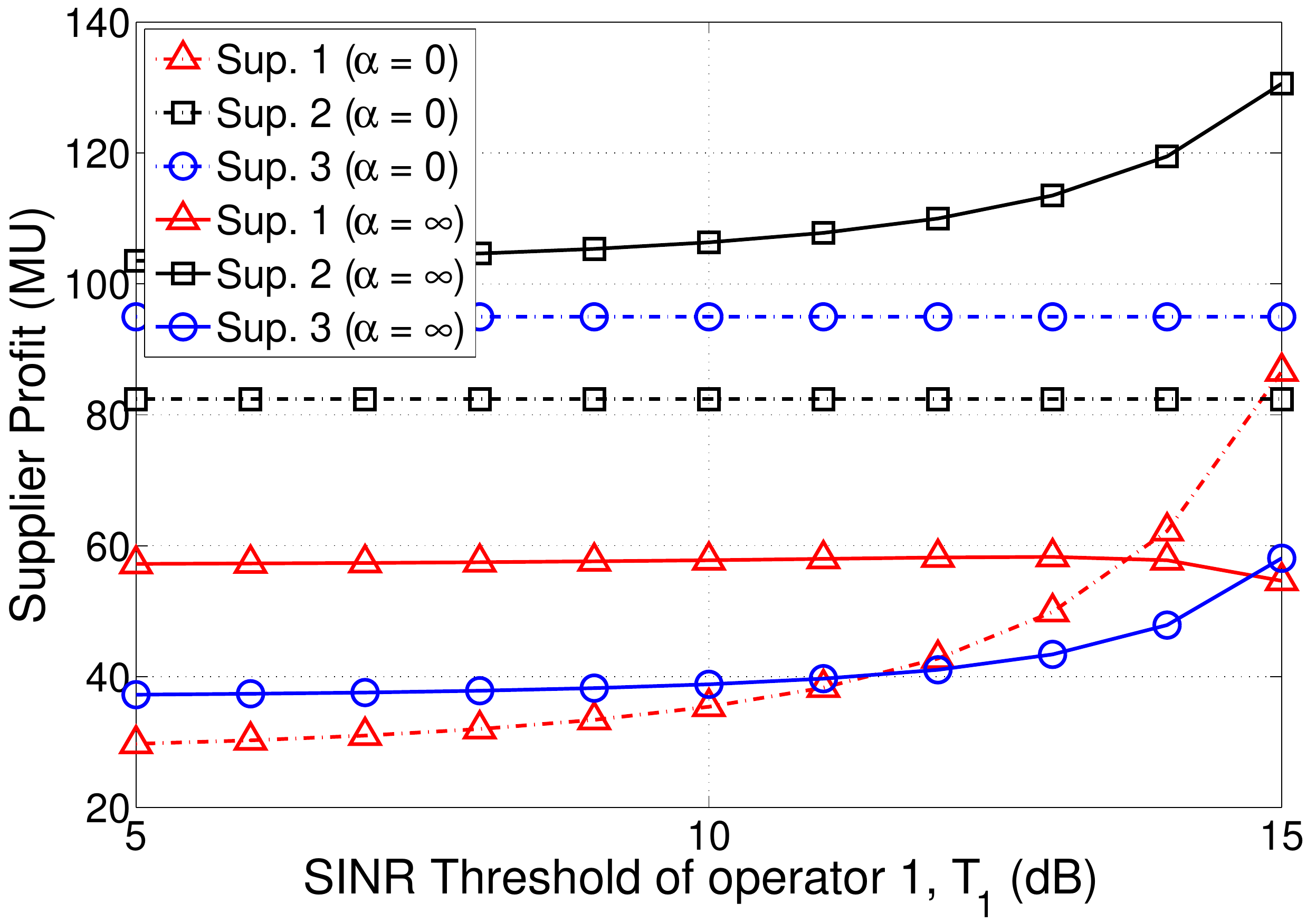}
  \vspace{-0.2cm}
  \caption{\small{Profit of energy suppliers versus SINR threshold of Op. 1.}}
  \label{profit_vs_T1}
\end{figure}

In the sequel, the price sensitivity of suppliers $\gamma_{n} = 0, \forall n = 1, \ldots, N_{\text{R}}$ and $\alpha=0.5$ is used unless otherwise stated. Fig.~\ref{profit_vs_T1} plots the profit of the suppliers against the SINR threshold of Op. 1 for extreme values of $\alpha$. It can be observed that for $\alpha = 0$, i.e., Sum utility case, all the additional share of profit is received by Sup. 1 since it has the lowest profit among the suppliers. However, for $\alpha = \infty$, i.e, Max-Min fair case, operator $2$ already receives a higher profit and the additional share of the profit is given to~Sup.~2~and Sup.~3.

\begin{figure}[t]
\begin{center}
\subfigure[Energy production by suppliers.]{\label{E_production_vs_cost}\includegraphics[width=2.35in]{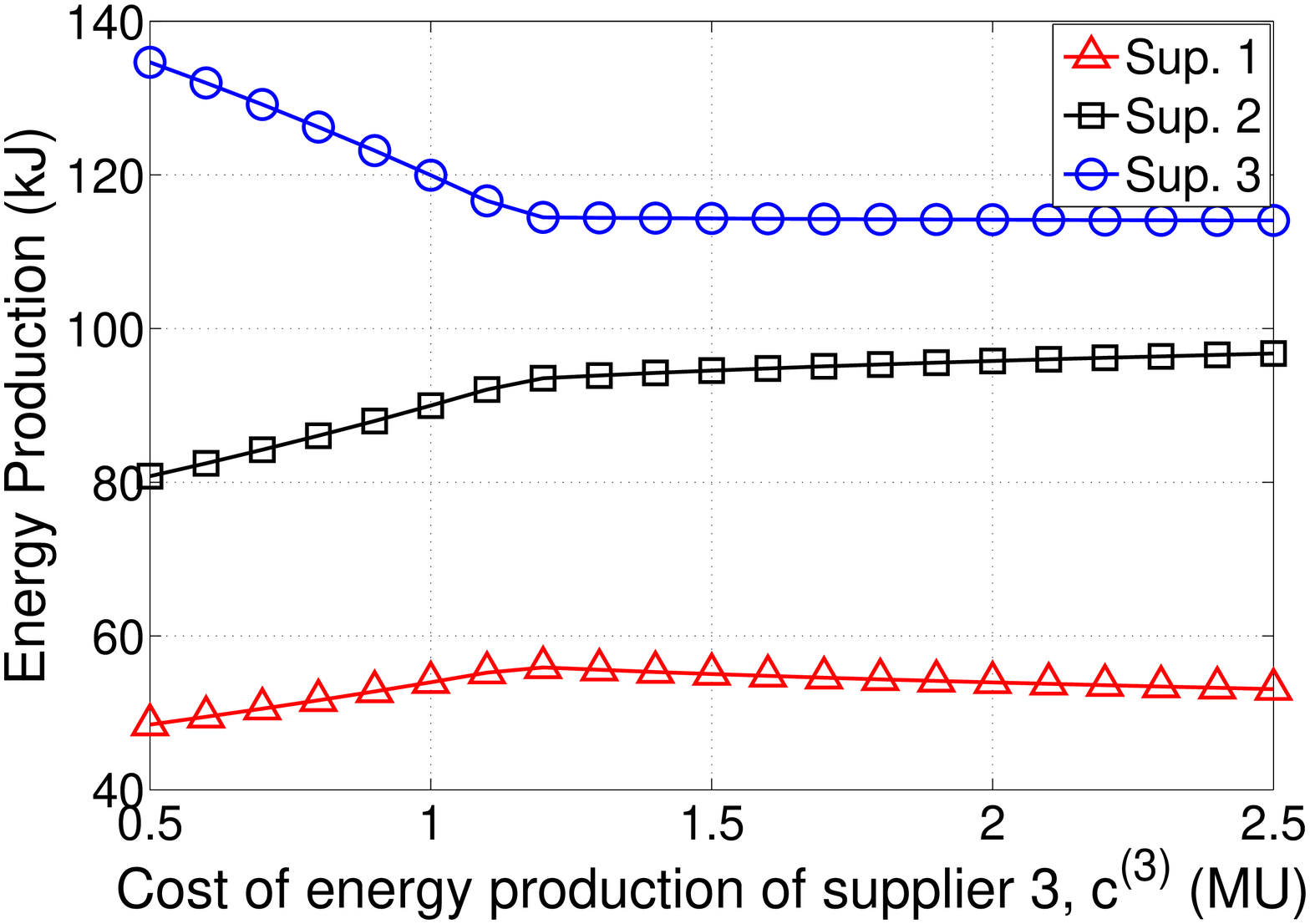}}\\
\vspace{-0.3cm}
\subfigure[Profit of energy suppliers.]{\label{profit_vs_cost}\includegraphics[width=2.35in]{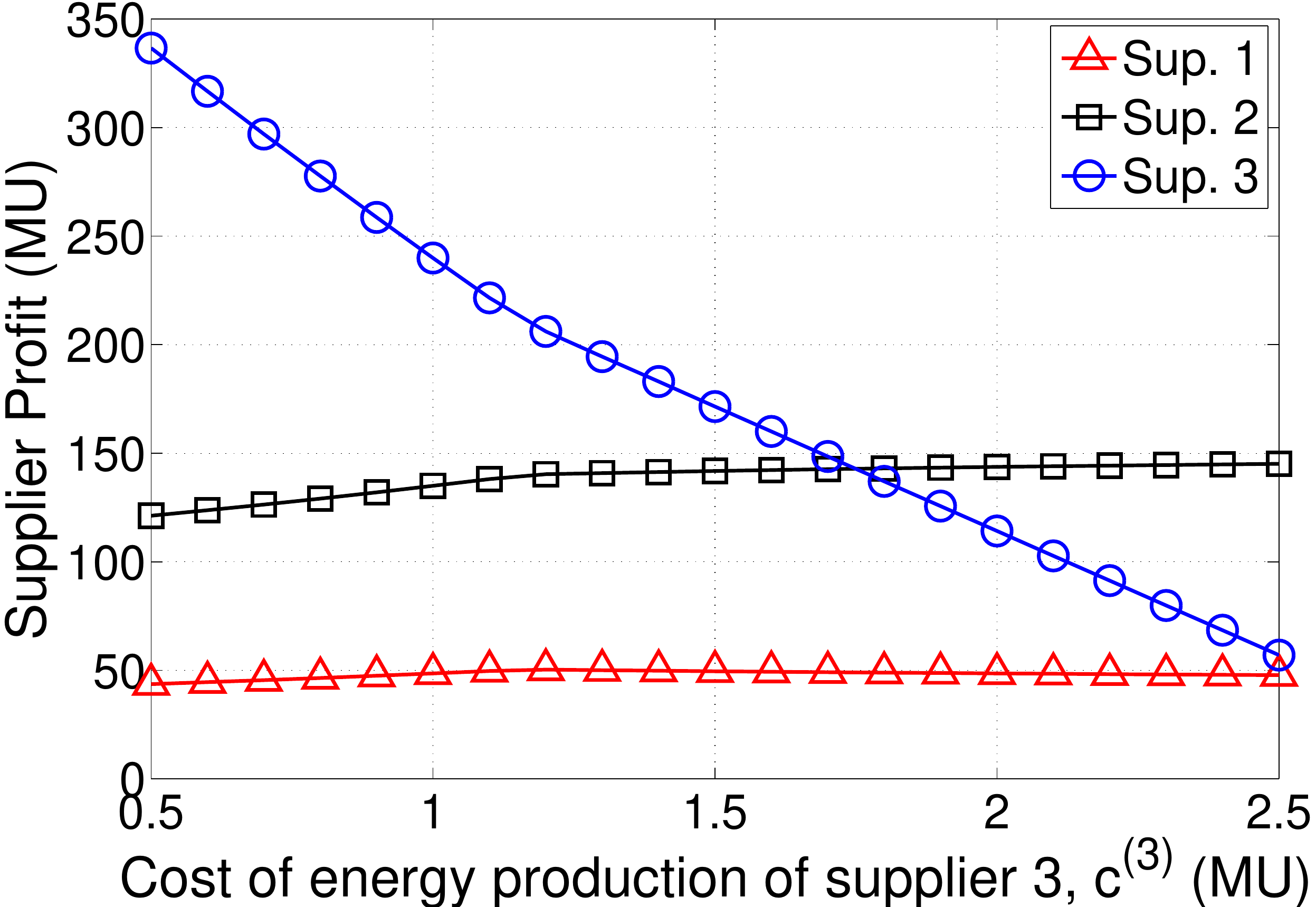}}
\end{center}
\vspace{-0.3cm}
\caption{\small{Energy production by suppliers and profit of suppliers versus production cost of Sup. 3 for $\alpha = 0.5$.} }
\label{fig3}  \vspace{-0.2cm}
\end{figure}

In Fig.~\ref{E_production_vs_cost}, the impact of increasing the cost of energy production of Sup. $3$ on the amount of energy produced by the suppliers is studied for $T_{1} = 15$ dB. It can be observed that increasing the cost of production reduces the profit margin of Sup. 3, thus leading to a reduction in its energy production. In order to meet the same energy demand, the production of Sup. 2 increases since Sup. 2 has the highest profit margin. An increase in the production of Sup. 1 is also observed since the fairness level $\alpha = 0.5$ does not allow any single supplier to exploit the demand. The corresponding profit of the suppliers is shown in Fig.~\ref{profit_vs_cost}. It is shown that the Sup. 3 profit is linearly falling as the cost increases while there is a gentle increase in Sup. 1 and Sup. 2 profits due to increased~production.

\begin{figure}[t]
  \centering
  \includegraphics[width=2.35in]{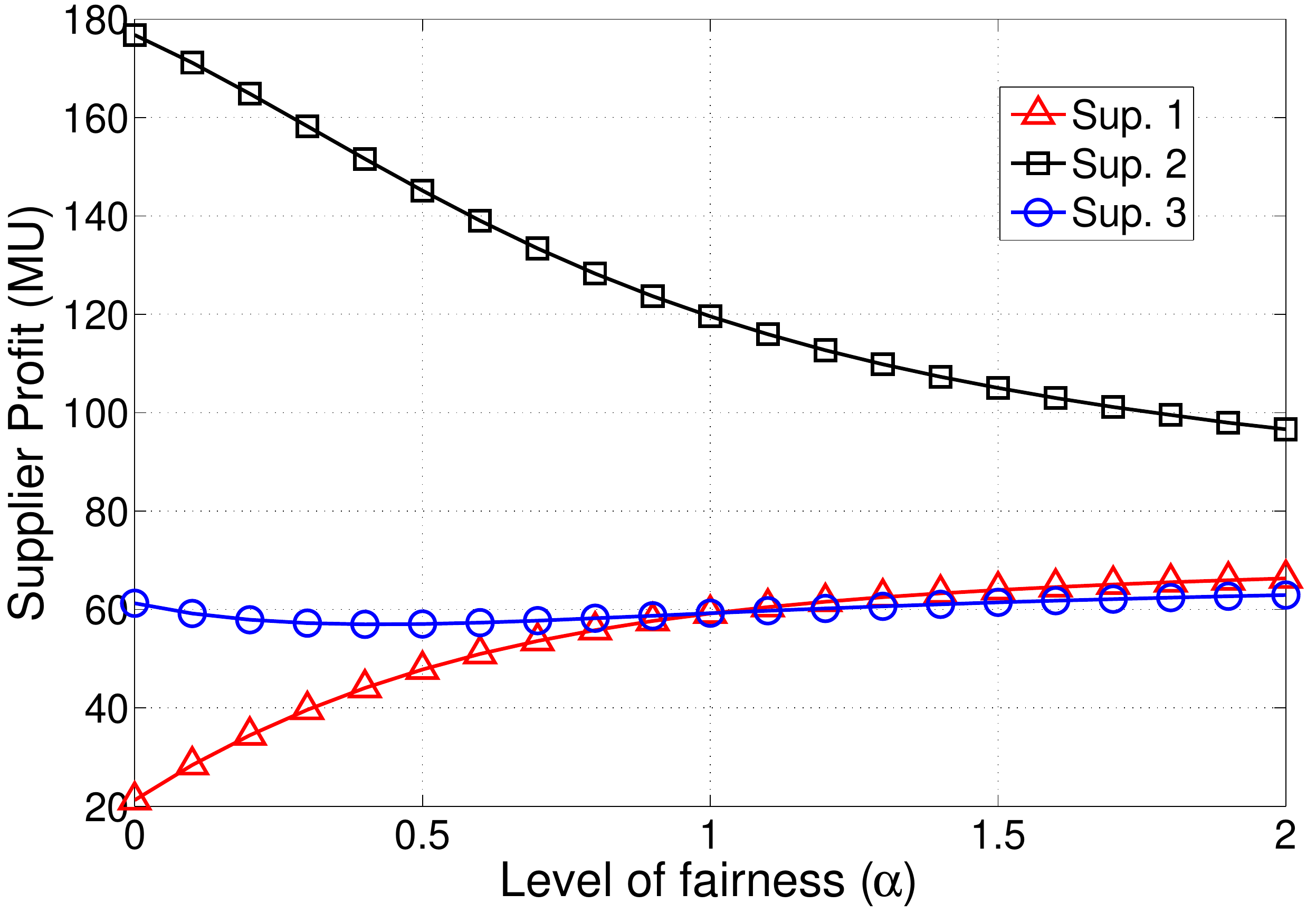}\vspace{-0.2cm}
  \caption{\small{Profit of energy suppliers versus the fairness level~$\alpha$.}}
  \label{profit_vs_alpha_fig}
\end{figure}

Finally, in Fig.~\ref{profit_vs_alpha_fig}, the effect of the fairness level $\alpha$ on the the profits of the suppliers is demonstrated. It is apparent that Sup. 2 achieves the highest profit due to its high profit margin followed by Sup. 3 and Sup. 1 in order of their profit margins. As $\alpha$ is increased, the distribution of profits among the three suppliers becomes fairer. Particularly, when $\alpha = 0$, the distribution of profit is biased. The supplier with the highest profit margin, i.e., Sup. 2 receives the highest share and the supplier with the lowest profit margin, i.e., Sup. 1 receives the lowest share. For $\alpha = 1$, i.e., Proportional fair, the distribution of profits is according to the profit margins, however, the discrepancy is smaller. As $\alpha$ is increased, the share of the profits equalizes further to increase fairness.

\vspace{-0.2cm}
\section{Conclusion}\label{sec5}
In this paper, a fair and optimized framework is proposed for energy production decisions made by suppliers in the smart grid powering cellular networks. Multiple energy suppliers are available in smart grid offering different energy prices and pollutant emissions. The goal is to first assess the power requirements of the mobile operators, for a given quality of transmission and the number of subscribers. Then, the framework needs to meet these requirements by intelligently providing energy from different suppliers in order to maximize the profitability of the suppliers. Stochastic geometry is used to quantify the network performance from the perspective of a typical user in terms of the average transmission power of the networks.
It is observed that using the proposed framework, suppliers adapts their energy production decisions under different situations such that the CO$_{2}$ emissions levels of the network do not exceed the set threshold and generate the optimal amount of energy for maximum profitability.

\vspace{-0.1in}
\bibliographystyle{ieeetr}
\bibliography{references}

\end{document}